\documentclass[a4paper,UKenglish]{article}

\usepackage[margin=2cm]{geometry}

\usepackage{microtype}

\usepackage{rotating}
\usepackage[T1]{fontenc}
\usepackage{wrapfig}
\usepackage{textcomp}
\usepackage{amstext}
\usepackage{graphicx}
\usepackage{booktabs}
\usepackage{rotating}
\usepackage{multirow}
\usepackage{hyperref}

\makeatletter

\providecommand{\\}{\\}

\usepackage{graphicx}

\makeatother

\usepackage{babel}
\begin{document}

\title{Intriguingly Simple and Efficient Time-Dependent Routing in Road
Networks\footnote{Partial support by DFG grant FOR-2083 and EU grant 288094 (eCOMPASS).}}

\author{Ben Strasser\\
Karlsruhe Institute of Technology (KIT), Karlsruhe, Germany\\
  \texttt{strasser@kit.edu}}

\maketitle
\begin{abstract}
We study the earliest arrival problem in road networks with static time-dependent functions as arc weights. 
We propose and evaluate the following simple algorithm: (1) average
the travel time in $k$ time windows, (2) compute a shortest time-independent
path within each window and mark the edges in these paths, and (3)
compute a shortest time-dependent path in the original graph restricted
to the marked edges. 
Our experimental evaluation shows that this simple algorithm
yields near optimal results on well-established benchmark instances.
We additionally demonstrate that the error can be further reduced
by additionally considering alternative routes at the expense of more marked edges. 
Finally, we show that the achieved subgraphs are small
enough to be able to efficiently implement profile queries using a
simple sampling-based approach. A highlight of our introduced algorithms is that they do not rely on linking and merging profile functions.
\end{abstract}

\section{Introduction}

Routing in road networks is an important and well-researched topic~\cite{bdgmpsww-rptn-14}. 
It comes in many varieties, nearly all of which start by formalizing the road network as a weighted directed graph.
Nodes correspond to positions in the network and edges correspond to road segments. 
The weights are the travel time a car needs to traverse a road segment. 
A common assumption is that these travel times are time-independent scalars that do not change throughout the day. 
With this assumption the problem boils down to the classical shortest path problem: Given a weighted graph, a source node $s$ and a target node $t$, find a shortest $st$-path.

A problem is that road networks are huge making even near-linear-running-time algorithms, such as Dijkstra's algorithm \cite{d-ntpcg-59}, too slow.
Therefore speedup based techniques were developed~\cite{bdgmpsww-rptn-14}: 
In a first slow preprocessing phase, auxiliary data is computed. 
In a second fast query phase, shortest path queries are answered in sublinear running time using this auxiliary data. 
The details of these techniques are not important for our paper as we only use these techniques as a black box. 
We use Contraction Hierarchies (CH)~\cite{gssv-erlrn-12} but they are exchangable with other techniques. 
See \cite{bdgmpsww-rptn-14} for an overview of other speedup based techniques.

Unfortunately the assumption that travel times are constant throughout
a day is too simplistic in practice. We cannot just ignore traffic
congestion, rush hours or bridge closures. This gives rise to the
earliest arrival problem~\cite{ch-tsrtn-66}. In addition to $s$ and $t$ we are given
a departure time $\tau$ at $s$. The problem now consists in computing
the fastest $st$-path for the departure time $\tau$. We say that
the earliest arrival problem is time-dependent, whereas the classic
shortest path problem is time-independent. Dijkstra's algorithm 
\cite{d-ntpcg-59} can be augmented to solve the problem~\cite{d-aassp-69} but it remains too slow.

The specifics of the earliest arrival problem vary with how the time-dependency is modelled. 
Through this paper we assume that edges are weighted by a travel time function. 
These functions map a moment in time when a car enters an edge onto the travel time needed at this moment.
We assume that we have full knowledge of these functions and that they are available during preprocessing as piece-wise linear functions. 
We further assume that these functions are periodic, i.e., the time-horizon is one day and time wraps around at midnight. 
This modeling is not new and has been studied extensively \cite{dw-tdrp-09,dbs-a-10,d-tdsr-11,ndls-bastd-12,dn-crdtd-12,bgsv-mtdtt-13,fhs-octds-14,bdpw-dtdrp-16,kmppwz-eotdr-16}.
One further assumes that a car that enters an edge earlier must leave the edge earlier than a car that entered at a later moment.
This property is called FIFO~\cite{or-spmda-90}.
Historic data is usually not available for every edge.
Many travel time functions are therefore constant. 
Were refer edges with a non-constant function as time-dependent edges.

This model excels at capturing recurrent traffic situations such as
daily rush hours. However, this setup does not enable us to model
sporadic events such as for example accidents or large unexpected
traffic jams. This is a different form of time-dependency, which is
usually handled using different speedup techniques which are capable
of quickly updating scalar edge weights. Such a weight update is
called customization. Examples of such techniques are Customizable
Contraction Hierarchies (CCH) \cite{dsw-cch-15} or Multi-Level extensions
of Dijkstra's algorithm (MLD/CRP) \cite{sww-daola-00,hsw-emlog-08,dgpw-crprn-13}. It is
non-trivial to combine these two worlds of time-dependency but works with this goal exist \cite{bdpw-dtdrp-16,dw-lbrdg-07}.
We introduce and evaluate a new simple algorithm called TD-S to handle the problem setting with travel time functions. 
We give ideas on how to extend it to the combined setting in the outlook.

The classical shortest path problem and the customizable techniques usually compute exact solutions, i.e., they find a shortest path and not some short but not necessarily shortest path. 
In the setting with weight functions, computing exact solutions has proven to be a lot more difficult. 
Fortunately, this is not necessary in practice as no two days will be exactly the same. 
Rush hours are shifted by several minutes in unpredictable ways. 
As a consequence the travel time functions have large error margins in practice and optimizing the path length below these margins is not useful. 
Most works therefore consider errors. 
The absolute error is the absolute value of the difference between an optimal path's length and the computed path length.
The relative error is the absolute error divided by an optimal path's length.

Besides the earliest arrival problem we also consider the profile problem.
Given $s$ and $t$ compute, instead of a single arrival time, a function $f$ that maps the departure time at $s$ onto the arrival time at $t$.
Formulated differently: Solve the earliest arrival problem for all departure times simultaneously.
In theory the complexity of $f$ when represented as piecewise linear function can be superpolynomial in the instance size \cite{fhs-octds-14}.
Fortunately, real-world instances are far from this theoretical worst case behavior.
Further, we observe that on real-world instances the number of paths that are optimal over a day is significantly smaller than the number of interpolation points of $f$.
Another observation is that on our data the number of paths decreases the more realistic test instances get.
On our instance with the highest percentage of time-dependent edges we observe the least number of paths.

\subsection{Related Work.}

The authors of \cite{ndls-bastd-12} observed that ALT \cite{gh-cspas-05} can be applied directly to the graph weighted with the lower bounds of the travel time functions. 
This yields the simple algorithm TD-ALT, but its running times are, as already with basic unidirectional ALT, comparatively modest. 
In \cite{dn-crdtd-12} the technique was extended to TD-CALT by first coarsening the graph and then applying TD-ALT to the core. 
In~\cite{d-tdsr-11} SHARC \cite{bd-sharc-09} (a combination of Arc-Flags \cite{l-aefea-04} with contraction) was extended to the time-dependent setting. 
In~\cite{bgsv-mtdtt-13} CHs were augmented with time-dependent weights. 
\cite{bdpw-dtdrp-16} did the same with MLD/CRP. 
Another technique is FLAT~\cite{kmppwz-eotdr-16}. Here the idea is to precompute the solutions from a set of landmarks to every node and during the query phase to route all sufficiently long paths through a landmark.
With the exception of \cite{ndls-bastd-12} and \cite{kmppwz-eotdr-16} all techniques coarsen the input graph using contraction. 
Doing this requires transforming and combining the function weights using two basic operations called the linking and merging of travel time functions \cite{dw-tdrp-09}.
Implementing these is non-trivial and all implementations we know
of fight with numerical stability issues. A design goal of our algorithm
was therefore to avoid this set of problems by avoiding link and merge operations.

\subsection{Outline.}

We start by describing a baseline that we call the freeflow heuristic, then build on it and introduce our algorithms TD-S and TD-S+A and the profile
extension TD-S+P, experimentally evaluate them, compare our results
with related work, and finally give an outlook on live-traffic integration.

\section{The Simplest TD Routing Algorithm: The Freeflow Heuristic}

The freeflow travel time along an edge is the travel time assuming that there is no congestion, i.e., the minimum travel time over the whole day.
The simplest heuristic for time-dependent routing known to us makes use of the freeflow travel times and works as following:
\begin{enumerate}
\item Compute a shortest time-independent path according to the freeflow travel times.
\item Compute the time-dependent travel time along this path.
\end{enumerate}
It can obviously be combined with any time-independent speedup technique which automatically translates into very good query running times.
Another huge strength is its simplicity.
The main problem with it is that the computed paths can potentially differ significantly from the optimal ones, i.e., the computed solutions can have a significant error.
In practice the achieved errors are far from what may happen in theory, however still noticeable and thus problematic.
Our proposed algorithm can be seen as extension of the freeflow heuristic that aims to reduce this error to negligible amounts.

\section{Our Algorithm: TD-S}

Our algorithm is preprocessing-based. The offline phase consists
of the following steps:
\begin{enumerate}
\item Fix a small set of time intervals called time windows. 
\item For each time window and edge compute the average travel time in that window.
\item Each time window induces a time-independent graph. Preprocess each
of these graphs using your favorite time-independent speedup technique.
\end{enumerate}
The query phase of our proposed algorithm consists of the following steps:
\begin{enumerate}
\item For each time window compute the shortest time-independent $st$-path
using the employed speedup technique and mark the edges in the path.
\item Run a basic time-dependent search on the subgraph of the time-dependent
input graph induced by all marked edges.
\end{enumerate}
We use Contraction Hierarchies (CH) \cite{gssv-erlrn-12} as our time-independent speedup technique.
We chose four time windows whose boundaries roughly reflect the beginning
and end of the two daily rush hours. The precise windows we use are
0:00 to 6:00, 7:00 to 9:00, 11:00 to 14:00, and 17:00 to 19:00. It
is clear that the proposed algorithm is fast. What is not clear is
that the computed paths have a small error. Most of our experimental
evaluation revolves around showing that this is the case for 
all benchmark instances that we have access to.
All of these have been used in previous publications and therefore form a set of 
well-established benchmark instances. We call our algorithm TD-S for
\emph{T}ime-\emph{D}ependent \emph{S}imple routing.

\subsection{Adding Alternatives: TD-S+A}

For time-independent routing, alternative routes have been proposed \cite{adgw-arrn-13}. 
These are short but not necessarily shortest $st$-paths that are sufficiently different from each other.
A reasonable expectation is that a congested road can be bypassed using an alternative route. 
This is the idea used in our second algorithm: TD-S+A.

Instead of computing one path per time window during the query phase, we compute several alternatives for each time window and restrict the time-dependent search to the union of all these paths.
More precisely we use a basic CH-based alternative path algorithm \cite{adgw-arrn-13}:
We run the forward and backward searches until all meeting nodes on a path whose length is at most 1.2 times the shortest path are found. 
For each meeting node we unpack the path and mark its edges.
In \cite{adgw-arrn-13} the resulting paths are filtered to eliminate strange alternatives or alternatives that are too similar to the primary route. 
We skip these filters and mark the edges in each path found regardless of whether it is a good alternative. 
Using this method we mark among other edges all good alternatives.
Marking more edges is find in our setting as this can not increase the error. 
Only the running time of TD-S+A is affected. 
Note that, while this alternative algorithm is simple to implement, other algorithms exist \cite{adgw-arrn-13,k-hdara-13,krs-eepma-13,ls-csarr-13,bdgs-argrn-11} that yield different and possibly better alternatives.

\subsection{Computing Profiles: TD-S+P}

Our profile approximation algorithm is sampling-based. 
During the query phase, we first mark the edges as with TD-S and then run the pruned time-dependent search several times for departure times at regular intervals. 
Our approximated profile interpolates between the sampled points linearly.
Our algorithm is fast because we only have to mark the edges once per query but we run the time-dependent search numerous times.
For a sampling rate of $10$min, our algorithm would mark the edges as for TD-S, then run the time-dependent extension of Dijkstra's algorithm restricted to the marked edges with the departure times 0:00, 0:10, 0:20$\ldots$23:50.
Denote by $a_1,a_2,a_3\ldots a_{144}$ are the computed arrival times.
The arrival time of for example the departure time 0:07 is then assumed to be $0.3 a_1+0.7 a_2$.

\begin{figure}
\vspace{-2em}
\begin{centering}
\includegraphics{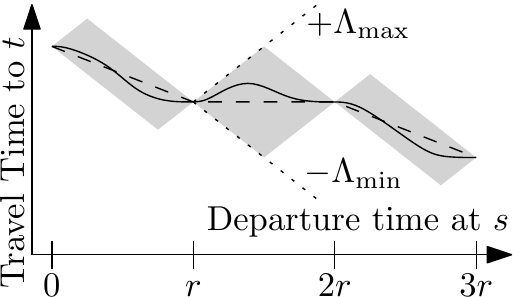}
\par\end{centering}

\caption{\label{fig:max-profile-error}The $st$-profile is the solid line.
It must be in the gray area. The dashed function is our approximation.}
\end{figure}%

We can bound the error by using an assumption also used in \cite{kmppwz-eotdr-16}:
Denote by $\Lambda_{\max}$ the maximum and by $-\Lambda_{\min}$
the minimum slope of every linear piece in every profile function and by $r$ the sample rate. Assuming that the sample
points are computed optimally, we can bound the absolute error. The
profile must be within the gray area depicted in Figure~\ref{fig:max-profile-error}.
As shown in the appendix this gives us a maximum absolute error of $r(\Lambda_{\max}+\Lambda_{\min})/4$.
In \cite{kmppwz-eotdr-16} the values $\Lambda_{\max}=0.19$ and $\Lambda_{\min}=0.15$
have been experimentally estimated for the Berlin instance. With $r=10$min
this gives a maximum error of $51$ seconds, which is good enough for all
but very local queries. Fortunately, local queries are also processed faster, which
allows us to increase the sample rate for these queries reducing the relative error.

Our approach is similar to the TRAP oracle from~\cite{kmppwz-eotdr-16} except that their approximation consists in picking the lower-envelope of trapez between two interpolation points instead of interpolating linearly. 

\section{Experimental Results}

\subsection{Setup}

\begin{table}
\begin{center}
\vspace{-3em}
\begin{tabular}{lrrrr}
\toprule
 & Nodes & Directed & TD-Edges  & avg. Break Points \\
 & {[}K{]} & Edges {[}K{]} & {[}\%{]} & per TD-Edge\\
\midrule
Ber & 443 & 988 & 20 - 28 & 65 - 74\\
Ger & 4688 & 10795 & 2 - 7 & 13 - 18\\
Eur & 18010 & 42189 & 0.1 - 6 & 11.1 - 11.2\\
\bottomrule
\end{tabular}
\end{center}

\caption{Node count, edge count, min. and max. percentage of time dependent
edges, and min. and max. number of break points per time-dependent
weight function for all instances. Min. and max. were computed over the various
days.}
\label{tab:instance-sizes}
\end{table}

We consider instances from three sources\footnote{We do not have access to further instances. Other common data sources, such as OpenStreetMap, do not include time-dependent data. For access to Germany \& Europe see \url{http://i11www.iti.uni-karlsruhe.de/resources/roadgraphs.php}.}.
The newest and presumably most accurate data was provided by TomTom in 2013. 
It contains the northern part of east Germany, i.e., Berlin, Brandenburg, and Mecklenburg-Vorpommern.
In the literature it is referred to by the largest city, i.e., Berlin. 
The data contains time-dependent weights for each day of the week. 
All Berlin experiments in the main part of the paper use Tuesday. 
Experiments for the remaining days are in the appendix.
The second data set was provided by PTV in 2006. 
It is based on NAVTEQ data and contains all of Germany. 
It contains 5 different types of days as Tuesday, Wednesday, and Thursday are assumed to be identical and are grouped as Ger-mid. 
The final instance is based on the time-independent DIMACS \cite{dgj-spndi-09} Europe graph which is also based on data provided by PTV. 
The time-dependent edge weights were synthetically generated by \cite{ndls-bastd-12}.
There are three variants denoted by L (low), M (medium), and H (high) where H has the most non-constant edge weights.
These instances are a superset of all instances evaluated in the related work \cite{dn-crdtd-12,bdpw-dtdrp-16,d-tdsr-11,kmppwz-eotdr-16,bdpw-dtdrp-16} and therefore form a well-established benchmark set. 
The sizes of the instances are reported in Table~\ref{tab:instance-sizes}.
All our experiments were run sequentially on a Xeon
E5-1630 v3 clocked at 3.70GHz with 128GB of 2133GHz DDR4 RAM.
For each of the three graphs we generated $10^{5}$ uniformly random
queries by picking a source and target node uniformly at random and
a source time picked uniformly at random over the whole day. All experiments
use the same set of random queries. We further generated random Dijkstra-rank
queries. The Dijkstra-rank of an $st$-query in a time-independent
graph is defined as follows: Order the nodes by increasing distance
from $s$. Let $p$ denote the position of $t$ in this order. The
Dijkstra-rank of the $st$-query is then $\left\lceil \log_{2}p\right\rceil $.
We generate $10^{4}$ queries per rank in the graph induced by the
edge travel time averaged over the whole day. 
The intuition is that a low Dijkstra-rank correlates with local queries
whereas a high Dijkstra-rank corresponds to long-distance queries.
Note that uniform random queries are with high probability long-distance queries. 
CH preprocessing needs per window about 6s on Berlin, about 100s on Germany, and about 1080s on Europe. 

\subsection{Random Uniform Queries}

\begin{table}
\begin{center}
\begin{tabular}{llrrrrrrr}
\toprule
  & & Opt. & \multicolumn{3}{c}{Relative Error} & Run. & \\
\cmidrule(lr){4-6}
 & & Solved & Avg. & Q99.9 & Max. & Time & speed\\
 & Algorihm & {[}\%{]} & {[}$\cdot10^{-7}${]} & {[}\%{]} & {[}\%{]} & {[}$ms${]} & up\\
\midrule
Ber-Tu & Freeflow & 76.4 & 12\,539.0 & 6.174 & 15.574 & 0.09 & 403\\
Ber-Tu & TD-S & 97.7 & 152.8 & 0.312 & 1.851 & 0.23 & 153\\
Ber-Tu & TD-S+A & 98.5 & 58.4 & 0.129 & 1.029 & 3.01 & 12\vspace{0.3em}\\

Ger-mid & Freeflow & 60.7 & 13\,996.4 & 4.730 & 12.372 & 0.24 & 2212\\
Ger-mid & TD-S & 95.0 & 261.4 & 0.383 & 2.024 & 0.60 & 894\\
Ger-mid & TD-S+A & 96.0 & 115.7 & 0.169 & 0.899 & 6.36 & 92\vspace{0.3em}\\

Eur-L & Freeflow & 10.4 & 77\,963.5 & 9.974 & 34.670 & 0.28 & 8539\\
Eur-L & TD-S & 41.5 & 9\,052.8 & 2.628 & 5.363 & 0.85 & 2701\\
Eur-L & TD-S+A & 53.2 & 2\,715.5 & 1.043 & 2.439 & 10.51 & 215\vspace{0.3em}\\

Eur-M & Freeflow & 10.6 & 62\,378.9 & 9.907 & 30.714 & 0.29 & 8342\\
Eur-M & TD-S & 35.6 & 6\,658.2 & 2.275 & 6.258 & 0.81 & 2857\\
Eur-M & TD-S+A & 41.3 & 3\,219.8 & 1.084 & 3.450 & 8.87 & 257\vspace{0.3em}\\

Eur-H & Freeflow & 11.8 & 94\,764.8 & 20.292 & 37.244 & 0.29 & 8413\\
Eur-H & TD-S & 35.3 & 7\,183.3 & 2.190 & 8.732 & 0.76 & 3126\\
Eur-H & TD-S+A & 40.2 & 3\,690.6 & 1.290 & 4.100 & 7.55 & 313\\
\bottomrule
\end{tabular}
\end{center}

\caption{\label{tab:uniform}Random uniform earliest arrival
queries for various algorithms.}
\end{table}

We ran a basic exact time-dependent extension of Dijkstra's algorithm that evaluates weight functions using a binary search as optimal baseline, the freeflow heuristic, TD-S, and TD-S+A on all uniform random queries.
The results are in Table \ref{tab:uniform} (and Tables~\ref{tab:uniform-freeflow}, ~\ref{tab:uniform-td-s} and~\ref{tab:uniform-td-s+a} of the appendix).
We report the number of queries answered optimally, several relative error measures (average, 99.9\%-quantile\footnote{The $\alpha$-quantile of a set $S$ is the smallest $x\in S$ such that $\alpha \cdot |S|$ elements from $S$ are smaller than $x$.}, and the maximum), the running time of TD-S and TD-S+A, and the speedup compared to the time-dependent variant of Dijkstra's algorithm.

We observe that our algorithms are always significantly faster than the exact baseline with TD-S having a speedup of up to 3\,126. 
Using alternatives routes costs about a factor 10 in terms of running time, whereas the decrease in relative error is lower. 
With the exception of a few outliers, all relative errors are very close to zero and are essentially negligible.
We observe that we need to discern between the real-world instances (Germany \& Berlin) and the synthetic instance (Europe).
The later is harder for TD-S as the errors are larger with respect to every metric. 
We explain why Europe is harder in detail when discussing profile query results.
On the real-world instances a large fraction of the queries are even solved optimally. 
The error quantiles are always below 0.38\%. 
The maximum values are below 2.0\%. 
This shows that no query was found where TD-S performed poorly. 
However, it also shows that there are a few outliers where it performs significantly poorer than on average. 
The freeflow heuristic is definitely worse in terms of errors than TD-S, but better than one would expect. 

\subsection{Dijkstra-Rank}

\newcommand{\maketable}[3]{
	\begin{figure}
		\begin{center}
		\includegraphics[width=0.43\textwidth]{{#1/accel.csv_correct_query_percentage}.png}
		\hspace{0.1em}
		\includegraphics[width=0.43\textwidth]{{#1/alt_accel.csv_correct_query_percentage}.png}
		\end{center}
		\vspace{-3.1em}
		\begin{center}
		\includegraphics[width=0.43\textwidth]{{#1/accel.csv_total_rel_error}.png}
		\hspace{0.1em}
		\includegraphics[width=0.43\textwidth]{{#1/alt_accel.csv_total_rel_error}.png}
		\end{center}
		\vspace{-3.1em}
		\begin{center}
		\includegraphics[width=0.43\textwidth]{{#1/accel.csv_total_abs_error}.png}
		\hspace{0.1em}
		\includegraphics[width=0.43\textwidth]{{#1/alt_accel.csv_total_abs_error}.png}
		\end{center}
		\caption{#2}
		\label{#3}
	\end{figure}
}

\newcommand{\makesinglecoltable}[3]{
	\begin{figure}
		\begin{center}
		\includegraphics[width=0.43\textwidth]{{#1/accel.csv_correct_query_percentage}.png}
		\end{center}
		\vspace{-3.1em}
		\begin{center}
		\includegraphics[width=0.43\textwidth]{{#1/accel.csv_total_rel_error}.png}
		\end{center}
		\vspace{-3.1em}
		\begin{center}
		\includegraphics[width=0.43\textwidth]{{#1/accel.csv_total_abs_error}.png}
		\end{center}
		\caption{#2}
		\label{#3}
	\end{figure}
}

\maketable{de/dido}{Dijkstra-rank results for Ger-mid. On the left side is TD-S and on the right side TD-S+A. \textbf{Warning}: The y-axis scales differ between left and right side.}{fig:dij-ger-rank}

We evaluate the performance of TD-S when answering local queries using Dijkstra-rank queries.
Figure~\ref{fig:dij-ger-rank} contains the results for Ger-mid. 
The results for some other instances are in the appendix.
We only report error related measures. 
The running times are as expected: Lower rank implies lower running time.

The top row of the figures shows that the number of correct queries
increases with decreasing rank. The remaining figures are boxplots.
However, the errors are always so small that all boxes stick to the
$x$-axis. Only outliers have a visible error. We can observe that
the maximum relative error can significantly increase with decreasing
rank. However, the absolute error does not. The explanation of this
effect is that queries with a low rank have a small travel time. Even
a small absolute error can seem large when compared to a small travel
time. Another observation is that, while alternatives have a limited
influence on average error values, they decrease the number of outliers
noticeably.

\subsection{Detailed Algorithm Evaluation}

\begin{table}
\begin{center}
\begin{tabular}{lllllrrrrr}
\toprule
\multirow{3}{*}{\begin{sideways}+A~~~\end{sideways}} & 
\multirow{3}{*}{\begin{sideways}0-6~~~\end{sideways}} & 
\multirow{3}{*}{\begin{sideways}7-9~~~\end{sideways}} & 
\multirow{3}{*}{\begin{sideways}11-14~~~\end{sideways}} & 
\multirow{3}{*}{\begin{sideways}17-19~~~\end{sideways}} & 
Opt. & \multicolumn{3}{c}{Relative Error} & \multicolumn{1}{l}{Run.}\\
\cmidrule(lr){7-9}
 & & & & & Sol. & Avg. & Q99.9 & \multicolumn{1}{l}{Max.} & \multicolumn{1}{l}{Time}\\
 & & & & & {[}\%{]} & {[}$\cdot10^{-7}${]} & {[}\%{]} & {[}\%{]} & {[}$ms${]}\\
\midrule
$\circ$ & $\bullet$ & $\circ$ & $\circ$ & $\circ$ & 62.7 & 11\,773.7 & 4.276 & 10.121 & 0.24\\
$\bullet$ & $\bullet$ & $\circ$ & $\circ$ & $\circ$ & 64.9 & 8\,332.7 & 2.921 & 7.524 & 1.99\\
$\circ$ & $\circ$ & $\bullet$ & $\circ$ & $\circ$ & 62.6 & 11\,148.5 & 3.796 & 12.126 & 0.24\\
$\circ$ & $\circ$ & $\circ$ & $\bullet$ & $\circ$ & 69.9 & 6\,803.2 & 2.704 & 8.730 & 0.24\\
$\circ$ & $\circ$ & $\circ$ & $\circ$ & $\bullet$ & 61.2 & 12\,795.4 & 3.991 & 12.711 & 0.24\\
$\circ$ & $\bullet$ & $\bullet$ & $\circ$ & $\circ$ & 89.0 & 1\,170.1 & 1.211 & 3.658 & 0.37\\
$\circ$ & $\bullet$ & $\circ$ & $\bullet$ & $\circ$ & 88.3 & 1\,398.9 & 1.442 & 6.176 & 0.36\\
$\circ$ & $\bullet$ & $\circ$ & $\circ$ & $\bullet$ & 90.7 & 742.9 & 0.825 & 6.176 & 0.37\\
$\circ$ & $\bullet$ & $\bullet$ & $\circ$ & $\bullet$ & 93.5 & 342.2 & 0.443 & 2.024 & 0.48\\
$\circ$ & $\bullet$ & $\bullet$ & $\bullet$ & $\bullet$ & 95.0 & 261.4 & 0.383 & 2.024 & 0.60\\
$\bullet$ & $\bullet$ & $\bullet$ & $\bullet$ & $\bullet$ & 96.0 & 115.7 & 0.169 & 0.899 & 6.39\\
\bottomrule
\end{tabular}
\end{center}

\caption{\label{tab:bucket-influence}Parameters influence on running time and error on Ger-mid.}

\end{table}

We have shown that TD-S as a whole works well. 
Next, we investigate which part contributes how much. 
In Table~\ref{tab:bucket-influence} we investigate the influence of the various time windows and of alternative routes. 
We observe that using a single time window already leads to a number of optimally solved queries of 61.2\% to 69.9\%. 
No time window performs particularly better than another.
However, we see that the impact of alternative routes is small even if only one window is used. 
Combining any window with the 0:00 to 6:00 window already yields 88.3\% to 90.7\% optimally solved queries. 
Adding further windows and alternative routes further increases this value. 

The running time reflects the configuration used. 
Activating a window adds $\approx$0.12ms per window. 
Using alternative routes increases the running time by about a factor 10.

\begin{table}
\begin{center}
\begin{tabular}{rcccr}
\toprule
\multirow{3}{*}{\begin{sideways}+A~~~~\end{sideways}}  & CH & CH & TD- & \\
 & Dist. & Path & Search & \#Marked \\
 &  [ms] &  [ms] &  [ms] &  Arcs\\
\midrule
$\circ$ & 0.30 & 0.21 & 0.20 & 1108\\
$\bullet$ & 0.34 & 3.33 & 3.33 & 20798\\
\bottomrule
\end{tabular}
\end{center}

\caption{\label{tab:running-time-parts}Running time of TD-S on Ger-mid.}
\end{table}%
We examine in further detail what part of our algorithm is responsible for the overall
running time and report the results in Table~\ref{tab:running-time-parts}.
A first observation is that the time needed to extract the paths and
to run the pruned time-dependent search are essentially the same.
We explain this effect by the fact that the workload for both tasks
is essentially proportional to the edges in the search graph. This
is consistent with the increase in search graph size and running time
when activating alternative routes: The number of arcs, the path unpacking
running time, and the pruned time-dependent search all increase by
about a factor 19. The time needed to explore the CH search spaces
and determine the meeting nodes increases significantly less. The
difference in running time between the Tables~\ref{tab:running-time-parts}
and~\ref{tab:bucket-influence} comes from measurement overhead and
the complete separation of the distance query and the path extraction,
which requires storing intermediate data to the main memory. In our
main implementation these are interleaved.

\subsection{Detailed Correctness Evaluation}

\begin{figure}
\begin{center}
\includegraphics[width=0.95\textwidth]{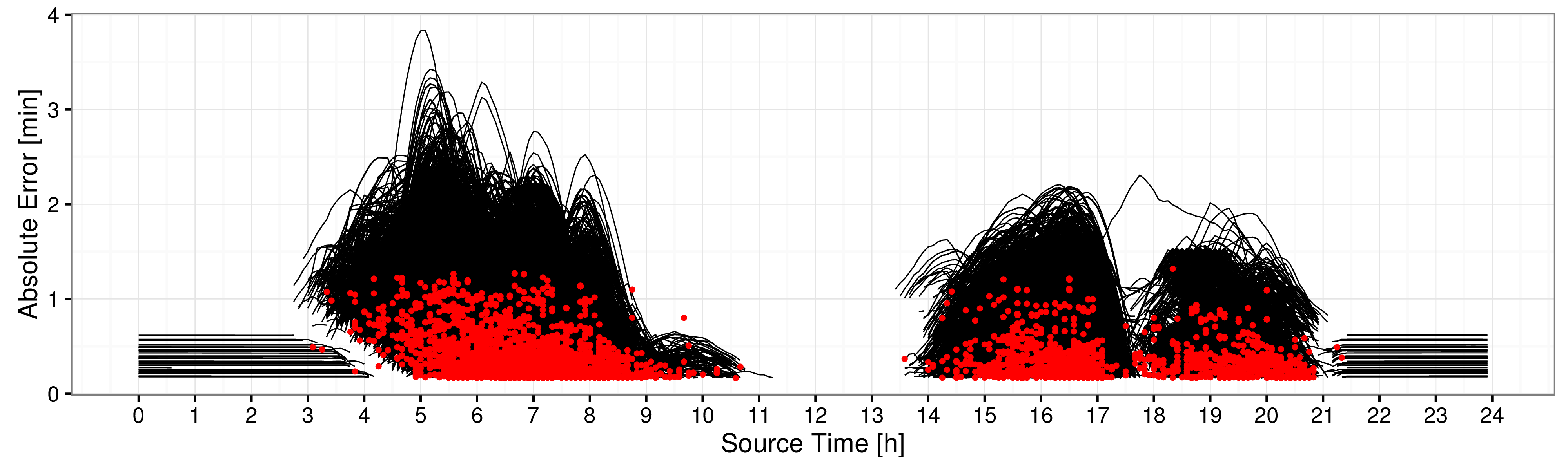}
\end{center}

\begin{center}
\vspace{-1.75em}
\includegraphics[width=0.95\textwidth]{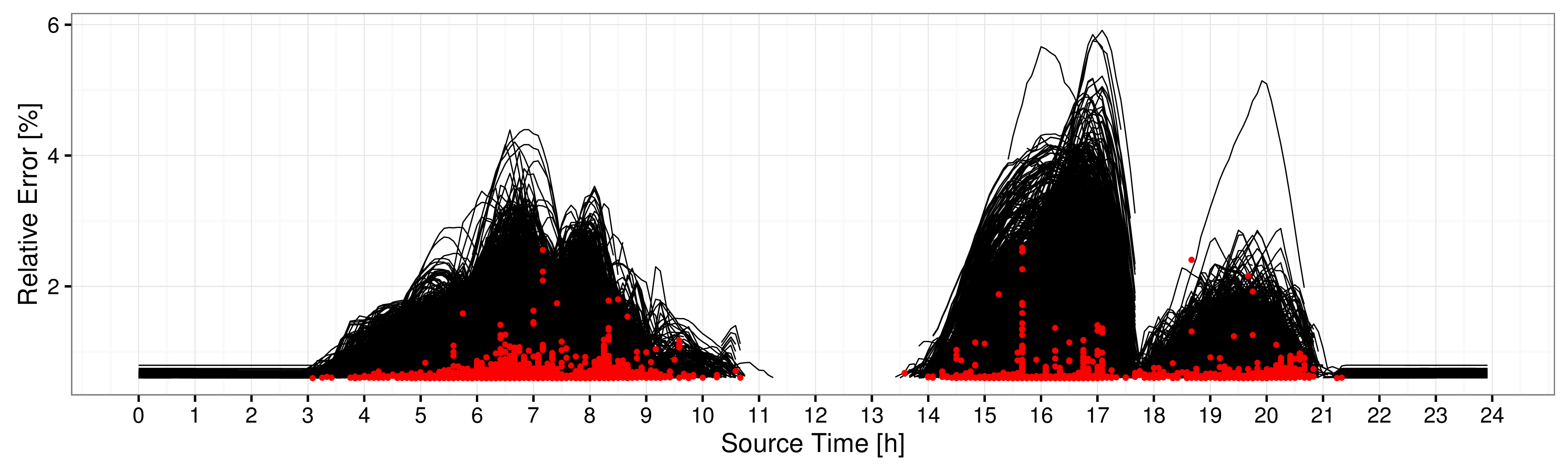}
\end{center}
\caption{\label{fig:extensive}Absolute (top) and relative (bottom) errors as function of source time for Berlin-Tu. Queries with the same source and target nodes and a source time difference equal to the sample rate are grouped and connected using lines. Queries that form their own group are represented as red dots.}
\end{figure}

In the previous section, we have reported absolute and relative errors for uniform random queries and Dijkstra-rank queries and have shown that TD-S behaves well with respect to these tests.
Unfortunately, we can not rule out that queries have escaped our sampling method for which TD-S behaves particularly bad.
Ideally, we would like to test every possible query. 
However, the number of queries is huge: We have to test every combination of source node, target node, and source time.
It is clear that we cannot test all queries because of limited computational resources.

We therefore try to get as close as possible: For the Berlin Tuesday instance, we consider the set $S$ of every 100-th node.
We compute paths between all nodes of $S$.
We sample the source time at a 5 min rate.
This gives us $5.7$ billion queries. 
During several days, we solved each of these queries using a 48-core machine using Dijkstra's algorithm and TD-S.
97.74\% of these queries were solved optimally by TD-S.
Another 2.23\% were solved with an absolute error below 10s or a relative error below 0.5\%, i.e., solved quasi optimally.
Only 0.03\% of all queries are left after filtering optimal and quasi optimal queries.

Figure~\ref{fig:extensive} depicts the absolute and relative errors of these 0.03\% queries in function of the source time. 
We group queries with the same source and target node together.
For every group with more than one query, we plot a black curve.
For every group with only one query, we plot a single red dot.

There are three bumps. 
The first bump corresponds to the morning rush hour whereas the other two correspond to the evening rush hour. 
The hole in the evening rush hour is explained by the 17:00-19:00 window.

There are 7 paths that are sub-optimal in the middle of the night. 
On these, the morning rush hour begins particularly early and therefore the 0:00-6:00 window misses the freeflow path, which would be optimal.

\subsection{Profiles}

\begin{table}
\begin{centering}
\begin{tabular}{lrrrrrr}
\toprule
 & Run. & \multicolumn{5}{c}{Path Count}\\
\cmidrule(lr){3-7}
 & {[}ms{]} & avg. & Q90\% & Q99\% & Q99.9\% & max.\\
\midrule
Ber-Tu & 4.1 & 1.4 & 2 & 4 & 7 & 12\\
Ger-mid & 8.0 & 1.5 & 2 & 4 & 7 & 13\\
Eur-L & 16.7 & 62.3 & 102 & 124 & 133 & 140\\
Eur-M & 15.0 & 36.9 & 68 & 92 & 107 & 123\\
Eur-H & 14.0 & 23.2 & 47 & 72 & 90 & 111\\
\bottomrule
\end{tabular}
\par\end{centering}

\caption{\label{tab:profile}Running time and distinct path count for TD-S+P. The sampling rate $r$ is 10min.}
\end{table}

We evaluated TD-S+P with a sample rate $r=10$min on uniform random queries and report the results in Table~\ref{tab:profile}. 
We first discuss the achieved running times. 
A 10min sampling rate implies 144 sample points over a whole
day. From Table~\ref{tab:running-time-parts} we know that running
the pruned time-dependent search needs about a third of the basic
algorithm's running time. For Eur-H we would therefore expect
a running time of $144\cdot0.76/3=36.48$ms. Yet our results report
only 14.0ms. The difference is a cache effect: The first
time-dependent search loads the subgraph into the cache. All consecutive
searches have nearly no cache misses. 

Besides computing the profile, we also counted how many paths are optimal
for at least one departure time. Our experiments show a very clear
difference between the real-world and the synthetic instances. In
the real-world instances only very few paths exist: On average there
are about 1.4 paths. Even the 99.9\% quantile is at most 8 paths.
This sharply contrasts with the synthetic instances: Here even the
average is at least 23.2 paths. The maximum goes up to 140. This is
only 4 paths less than the maximum of 144 paths discernable by our sampling
method. Another interesting observation is that Eur-L has \emph{more}
paths than Eur-H even though it has \emph{less} time-dependent edges.

\begin{figure}
\begin{centering}
\includegraphics{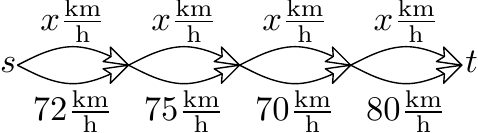}
\par\end{centering}

\caption{\label{fig:example}Abstract motorway (top) with neighboring rural roads (bottom).}
\end{figure}%

To explain this effect consider the instance depicted in Figure~\ref{fig:example}.
Suppose that all edges have the same length but the speeds vary. The
bottom edges have constant, i.e., time-independent speeds, whereas
the top edges have variable, i.e., time-dependent, speeds. We assume
for simplicity that all time-dependent edges have the same speed $x$.
This setup occurs in the generated instances: Suppose that the top
edges form a motorway, whereas the bottom edges are rural roads, slip road, or
even just service areas. According to~\cite{ndls-bastd-12} time-dependency
was only generated for edges that are important according to a Highway-Hierachy,
i.e., only motorways. If $x$ varies between $70$km/h
and $130$km/h, which is a reasonable assumption for a motorway that
has congestion in rush hours, then already this simple example generates
5 paths. It remains to argue why this does not occur in the real-world
instances. We believe that two effects are at work: First, people
are very good at congesting urban regions uniformly. As a result the
bottom edges would also be time-dependent in the real-world.
Secondly, the raw data used to generate both real-world instances
is not very fine-grained. For example the input data of the Germany
instance has speeds rounded to 5km/h steps. This is reasonable considering
that no two traffic days are the same and a fluctuation
of 5 km/h is to be expected. However, it also has the following effect:
If $x$ is always rounded to a 5km/h boundary then the path using
the 72km/h edge but not the 70km/h edge is never optimal. 
 Our sampling-based approach
in some situation hits the 72km/h edge or the time-dependent edge
above it but not both. At certain moments in time our algorithm therefore
choses the wrong edge. However, this only occurs when $x$ is close
to 72km/h, i.e., when the induced error is small. 
This observation
explains why comparatively few paths on the Europe instances
are optimal but nearly all are nearly optimal.
We can now also explain why Eur-H has less paths than Eur-L: More edges have time-dependency
and therefore the effects explained above happen less frequently.

\subsection{Comparison with Related Work}

\begin{table}[ht]
\begin{center}
\begin{tabular}{ccccrrrrr}
\toprule
 &  &  & \begin{sideways}\hspace{-1.9em}L\&M?\end{sideways} & \multicolumn{3}{r}{Relative Error {[}\%{]}} & \multicolumn{2}{c}{Run T. [ms] }\\
\cmidrule(lr){5-7} \cmidrule(lr){8-9}
 & &  &  & \multicolumn{1}{r}{avg.} & \multicolumn{1}{r}{Q99.9} & \multicolumn{1}{r}{max.} & ori & scaled\\
\midrule

TDCALT-K1.00 & \cite{dn-crdtd-12} & Ger-mid & $\bullet$ & $0$ & $0$ & $0$ & 5.36 & 3.77\\
TDCALT-K1.15 & \cite{dn-crdtd-12} & Ger-mid & $\bullet$ & 0.051 & n/r & $13.84$ & 1.87 & 1.31 \\
eco L-SHARC & \cite{d-tdsr-11} & Ger-mid & $\bullet$ & $0$ & $0$ & $0$ & 6.31 & 4.43 \\
heu L-SHARC & \cite{d-tdsr-11} & Ger-mid & $\bullet$ & n/r & n/r & 0.61 & 0.38& 0.27 \\
TD-CH & \cite{bgsv-mtdtt-13} & Ger-mid & $\bullet$ & $0$ & $0$ & $0$ & 0.75 & 0.54 \\
TDCRP (0.1) & \cite{bdpw-dtdrp-16} & Ger-mid & $\bullet$ & 0.05 & n/r & 0.25 & 1.92 & 1.38 \\
TDCRP (1.0) & \cite{bdpw-dtdrp-16} & Ger-mid & $\bullet$ & 0.68 & n/r & 2.85 & 1.66 & 1.19 \\

Freeflow & This Paper & Ger-mid & $\circ$ & 0.14 & 4.730 & 12.372 & 0.24 & 0.24\\

FLAT-$SR_{2000}$ & \cite{kmppwz-eotdr-16} & Ger-mid & $\circ$ & $1.444$ & n/r & n/r & 1.28 & 1.18\\

TD-S & This Paper & Ger-mid & $\circ$ & 0.003 & 0.383 & 2.024 & 0.60 & 0.60\\

TD-S+A & This Paper & Ger-mid & $\circ$ & 0.001 & 0.169 & 0.899 & 6.36 & 6.36\vspace{0.5em}\\

TDCALT-K1.00 & \cite{dn-crdtd-12} & Eur-H & $\bullet$ & $0$ & n/r & $0$ & 121.4\phantom{0} & 85.31\\
TDCALT-K1.25 & \cite{dn-crdtd-12} & Eur-H & $\bullet$ & 0.549 & n/r & $15.52{}^{\dagger}$ & 5.4\phantom{0} & 3.79\\
eco L-SHARC & \cite{d-tdsr-11} & Eur-H & $\bullet$ & $0$ & n/r & $0$ & 38.29 & 26.09 \\
heu L-SHARC & \cite{d-tdsr-11} & Eur-H & $\bullet$ & n/r & n/r & 1.60 & 2.13 & 1.50\\
TD-CH & \cite{bgsv-mtdtt-13} & Eur-H & $\bullet$ & $0$ & $0$ & $0$ & 2.11 & 1.52\\
TDCRP (0.1) & \cite{bdpw-dtdrp-16} & Eur-H & $\bullet$ & 0.04  & n/r & 0.29 & 6.47 & 4.64 \\
TDCRP (1.0) & \cite{bdpw-dtdrp-16} & Eur-H & $\bullet$ & 0.54  & n/r & 3.21 & 5.75 & 4.13 \\

Freeflow & This Paper & Eur-H & $\circ$ & 0.948 & 20.292 & 37.244 & 0.29 & 0.29\\
TD-S & This Paper & Eur-H & $\circ$ & 0.072 & 2.190 & 8.732 & 0.76 & 0.76\\
TD-S+A & This Paper & Eur-H & $\circ$ & 0.037 & 1.290 & 4.100 & 7.55 & 7.55\\
\bottomrule 
\end{tabular}
\end{center}

\caption{\label{tab:comparison}Comparison with related work for uniform random queries. 
The references point to the publication from which the numbers were taken.
``n/r'' stands for not reported. 
Running times are reported unscaled and scaled by processor clock speed. ``L\&M?'' indicates whether a technique needs linking and merging.
}
\end{table}

In Table~\ref{tab:comparison} we compare TD-S with other algorithms on Ger-mid and Eur-H.
The appendix contains a comparison for Berlin-Tu.

Comparing error values is difficult. 
The state-of-the-art consists in reporting the maximum relative error over $10^x$ uniform random queries where $x$ varies among papers. 
Unfortunately, the maximum error heavily depends on $x$: The more queries are performed, the larger the maximum error usually gets.
Further, uniform random queries are heavily biased towards long-distance queries.
This setup can therefore hide the worst queries as can be seen from Figure~\ref{fig:extensive}.
This is not a problem exclusive to TD-S but one inherent to the experimental setup.
However, as we cannot rerun the experiments of every related technique, we follow the established setup to get the fairest comparison with related work possible.

TD-S is one of the few algorithms that do not need link and merge operations and thus does not inherit all of the associated problems. 
Among these TD-S clearly dominates the competition in terms of error.
The Freeflow heuristic only wins, for obvious reasons, in terms of query running time.
Compared with link- and merge-based techniques TD-S is still highly competitive but there are a few techniques, such as TD-CH and heu L-SHARC, which manage to achieve smaller errors with smaller or comparable running times.
This advantage comes at the price of significantly more complex algorithms.
Further, for nearly all practical applications a running time below 1ms is good enough and a 1\% error is small compared to the error inherent to the input data. 
For example the input speeds of the Germany instance are multiples of 5km/h. 
Consider a route of 100km driven at either 60km/h or 65km/h. 
The travel times are 100min or 92.3min receptively, which is a larger difference than 1\%.
In practice, the techniques are thus equivalent in terms of performance.
However, they are not equivalent in terms of implementation complexity. 
Here TD-S clearly wins.

\section{Outlook}

The basic TD-S algorithm solves the time-dependent earliest arrival
problem with static weight functions. This does not allow for live-traffic
updates. However, we believe that integrating these is simple: Besides
the four static time window graphs we propose to use a fifth graph that
reflects the current traffic situation. For this fifth graph we cannot
use a CH as integrating live-traffic would be too slow but a CCH \cite{dsw-cch-15} or
MLD/CRP \cite{sww-daola-00,hsw-emlog-08,dgpw-crprn-13} should work. TD-S would in that setting mark the paths from
all 5 graphs and run a pruned time-dependent search that is aware
of the current traffic situation. If we can get access to realistic
live-traffic events, we would like to evaluate this extension.

Profile queries have proven to be a more difficult problem setting than earliest arrival time.
Indeed, besides TD-S only TD-CH has been shown to be able to compute profiles on the large Europe instances in a reasonable query running time.
Fortunately, the approach employed by TD-S+P is not only applicable to TD-S. 
Combinations with for example SHARC or TD-CH are imaginable.
Investigating these combinations could be fruitful.

TD-S+A uses an elementary algorithm to compute alternatives. 
Our results show that, while there is a decrease in error rates, the overall improvements over TD-S are very modest.
This does however not mean that using other algorithms to compute alternatives such as \cite{k-hdara-13,krs-eepma-13,ls-csarr-13,bdgs-argrn-11} do not yield better results.
Further research could therefore prove useful.

\section{Conclusion}

We introduced TD-S, a simple and efficient routing algorithm that addresses the time-dependent routing problem with static travel-time functions.
We combined it with alternative routes yielding TD-S+A and extended it to an efficient profile algorithm TD-S+P.
A highlight of our algorithms is that they do not use profile link nor profile merge operations.

\textbf{A word of Caution:}
The good quality of TD-S's found paths depends on the structure of the instances it is run on. 
All data available to us indicates that real-world instances have this structure. 
However, the data at our disposition is not the best: The Germany instance is 10 years old, the Berlin instance is small, and the Europe instance is synthetic.
We cannot exclude that other realistic instances exist, where TD-S performs poorly.
However, as we tested a superset of all instances also evaluated by the competitor algorithms, it would in that case also be unclear how the competitors would perform on such a hypothetical instance.
We have therefore either shown that TD-S is a simple, highly competitive algorithm or that better benchmark instances are necessary and existing techniques should ideally be reevaluated on them. 
We believe that the former is the case but cannot rule out the later.
In either case we believe that our results form an important contribution to the research on this subject.

\textbf{Acknowledgments:}
I thank Julian Dibbelt for converting all instances into a uniform format and Julian Dibbelt and Michael Hamann for fruitful discussions.

\newpage
\appendix

\section{Profile Error Guarantee}

\begin{wrapfigure}{o}{0.5\columnwidth}%
\begin{centering}
\includegraphics{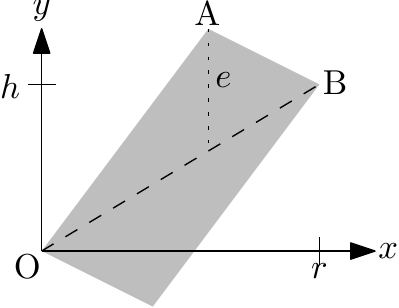}
\par\end{centering}

\caption{Proof Illustration.}
\label{fig:bound-proof}\end{wrapfigure}%

Consider the situation depicted in Figure~\ref{fig:bound-proof}.
Our objective is to compute the maximum vertical distance from the
dashed line inside of the gray region. As the triangle below the dashed
line is the same as the triangle above it rotated by 180\textdegree{},
we can focus solely on the upper triangle $OAB$. As the distance
grows from $O$ to $A$ and decreases from $A$ to $B$ we know that
the distance is maximum at $x=A_{x}$, i.e., the maximum vertical
distance is the length $e$ of the dotted segment. All coordinates
are constant except for the variable height $h$ of $B$. From the
application we know that $B_{x}=r$, and that the slope of the line
$OA$ is $\Lambda_{\max}$, and that the slope of $AB$ is $-\Lambda_{\min}$.
Our objective is to compute the maximum value of $e$ over all values
of $h$. We start by computing $e$ and then maximize the resulting
expression over $h$.

The line $OA$ is described by $y=\Lambda_{\max}x$, and the line
$OB$ is described by $y=\frac{h}{r}x$, and the line $AB$ is described
by $y=-\Lambda_{\min}x+(r\Lambda_{\text{\ensuremath{\min}}}+h)$.
By intersecting $OA$ and $AB$ we get 
\[
\Lambda_{\max}A_{x}=-\Lambda_{\min}A_{x}+(r\Lambda_{\text{\ensuremath{\min}}}+h)
\]
which can be solved for $A_{x}$ yielding

\[
A_{x}=\frac{r\Lambda_{\text{\ensuremath{\min}}}+h}{\Lambda_{\max}+\Lambda_{\min}}
\]
 which leads to 
\begin{eqnarray*}
e & = & \Lambda_{\max}A_{x}-\frac{h}{r}A_{x}\\
 & = & (\Lambda_{\max}-\frac{h}{r})A_{x}\\
 & = & \frac{(\Lambda_{\max}-\frac{h}{r})(r\Lambda_{\text{\ensuremath{\min}}}+h)}{\Lambda_{\max}+\Lambda_{\min}}
\end{eqnarray*}
which is an expression for the desired vertical height. As $0<\Lambda_{\max}$
and $0<\Lambda_{\min}$ the value of $e$ is maximum if and only if
\begin{eqnarray*}
 &  & (\Lambda_{\max}-\frac{h}{r})(r\Lambda_{\text{\ensuremath{\min}}}+h)\\
 & = & -\frac{h^{2}}{r}+(\Lambda_{\max}-\Lambda_{\text{\ensuremath{\min}}})h+r\Lambda_{\text{\ensuremath{\min}}}\Lambda_{\max}
\end{eqnarray*}
is maximum. As $-h^2 < 0$ this parabola is maximum when its derivative is zero.
We therefore compute the derivative
\[
-\frac{2h}{r}+(\Lambda_{\max}-\Lambda_{\text{\ensuremath{\min}}})
\]
which is zero for 
\[
h=\frac{r(\Lambda_{\max}-\Lambda_{\text{\ensuremath{\min}}})}{2}
\]
which we can insert into the expression of $e$ to obtain
\begin{eqnarray*}
\max_{h}e & = & \frac{(\Lambda_{\max}-\frac{r(\Lambda_{\max}-\Lambda_{\text{\ensuremath{\min}}})}{2r})(r\Lambda_{\text{\ensuremath{\min}}}+\frac{r(\Lambda_{\max}-\Lambda_{\text{\ensuremath{\min}}})}{2})}{\Lambda_{\max}+\Lambda_{\min}}\\
 & = & \frac{(\frac{\Lambda_{\max}+\Lambda_{\text{\ensuremath{\min}}}}{2})(\frac{r(\Lambda_{\max}+\Lambda_{\min})}{2})}{\Lambda_{\max}+\Lambda_{\min}}\\
 & = & \frac{r(\Lambda_{\max}+\Lambda_{\min})}{4}
\end{eqnarray*}
which was the absolute error bound we needed to compute.

\section{Full Tables}

Tables \ref{tab:uniform-freeflow}, \ref{tab:uniform-td-s}, and \ref{tab:uniform-td-s+a} contain the results for Freeflow, TD-S, and TD-S+A for all instances. Table \ref{tab:profile} is the extended version of Table \ref{tab:profile}.

\begin{table}
\begin{center}
\begin{tabular}{lrrrrrrr}
\toprule
  & Opt. & \multicolumn{3}{c}{Relative Error} & Run. & \\
\cmidrule(lr){3-5}
  & Solved & Avg. & Q99.9 & Max. & Time & speed\\
 & {[}\%{]} & {[}$\cdot10^{-7}${]} & {[}\%{]} & {[}\%{]} & {[}$ms${]} & up\\
\midrule

Ber-Mo & 75.9 & 14\,437.5 & 6.560 & 21.128 & 0.09 & 408 \\
Ber-Tu & 76.4 & 12\,539.0 & 6.174 & 15.574 & 0.09 & 403 \\
Ber-We & 76.2 & 12\,528.6 & 6.006 & 18.549 & 0.09 & 404\\
Ber-Th & 75.2 & 14\,147.7 & 6.515 & 17.822 & 0.09 & 405\\
Ber-Fr & 76.0 & 14\,998.2 & 6.649 & 22.565 & 0.09 & 407\\
Ber-Sa & 82.6 & 5\,749.9 & 3.962 & 11.408 & 0.08 & 412\\
Ber-Su & 84.5 & 3\,595.7 & 2.695 & 13.182 & 0.08 & 414\vspace{0.5em}\\
Ger-Mo & 60.9 & 13\,304.8 & 4.509 & 12.499 & 0.24 & 2318\\
Ger-mid & 60.7 & 13\,996.4 & 4.730 & 12.372 & 0.24 & 2212\\
Ger-Fr & 61.3 & 13\,480.1 & 4.341 & 11.900 & 0.24 & 2298\\
Ger-Sa & 72.8 & 6\,609.3 & 2.901 & 9.879 & 0.22 & 2498\\
Ger-Su & 77.7 & 4\,186.9 & 2.199 & 8.788 & 0.22 & 2575\vspace{0.5em}\\
Eur-L & 10.4 & 77\,963.5 & 9.974 & 34.670 & 0.28 & 8539\\
Eur-M & 10.6 & 62\,378.9 & 9.907 & 30.714 & 0.29 & 8342\\
Eur-H & 11.8 & 94\,764.8 & 20.292 & 37.244 & 0.29 & 8413\\
\bottomrule
\end{tabular}
\end{center}

\caption{\label{tab:uniform-freeflow}Random uniform earliest arrival
queries for the freeflow heuristic.}
\end{table}

\begin{table}
\begin{center}
\vspace{-2em}
\begin{tabular}{lrrrrrr}
\toprule
 &  Opt. & \multicolumn{3}{c}{Relative Error} & Run. & \\
\cmidrule(lr){3-5}
 & Solved & Avg. & Q99.9 & Max. & Time & speed\\

 & {[}\%{]} & {[}$\cdot10^{-7}${]} & {[}\%{]} & {[}\%{]} & {[}$ms${]} & up\\
\midrule

Ber-Mo & 97.7 & 164.5 & 0.315 & 2.562 & 0.23 & 152 \\
Ber-Tu & 97.7 & 152.8 & 0.312 & 1.851 & 0.23 & 153 \\
Ber-We & 97.8 & 158.8 & 0.330 & 2.074 & 0.23 & 155 \\
Ber-Th & 97.7 & 162.2 & 0.318 & 1.402 & 0.23 & 152 \\
Ber-Fr & 97.7 & 180.9 & 0.351 & 2.739 & 0.23 & 155 \\
Ber-Sa & 98.7 & 46.7 & 0.108 & 2.316 & 0.22 & 151 \\
Ber-Su & 99.0 & 30.2 & 0.077 & 0.793 & 0.21 & 152 \vspace{0.5em}\\
Ger-Mo & 94.9 & 248.8 & 0.389 & 2.648 & 0.61 & 909 \\
Ger-mid & 95.0 & 261.4 & 0.383 & 2.024 & 0.60 & 894 \\
Ger-Fr & 94.1 & 291.4 & 0.391 & 1.617 & 0.59 & 920 \\
Ger-Sa & 97.7 & 72.2 & 0.143 & 0.991 & 0.56 & 974 \\
Ger-Su & 98.4 & 28.4 & 0.066 & 0.513 & 0.55 & 995 \vspace{0.5em}\\
Eur-L & 41.5 & 9052.8 & 2.628 & 5.363 & 0.85 & 2701 \\
Eur-M & 35.6 & 6658.2 & 2.275 & 6.258 & 0.81 & 2857 \\
Eur-H & 35.3 & 7183.3 & 2.190 & 8.732 & 0.76 & 3126 \\
\bottomrule
\end{tabular}
\end{center}

\caption{\label{tab:uniform-td-s}Random uniform earliest arrival
queries for TD-S.}
\end{table}

\begin{table}
\begin{center}
\begin{tabular}{lrrrrrrr}
\toprule
  & Opt. & \multicolumn{3}{c}{Relative Error} & Run. & \\
\cmidrule(lr){3-5}
  & Solved & Avg. & Q99.9 & Max. & Time & speed\\
 & {[}\%{]} & {[}$\cdot10^{-7}${]} & {[}\%{]} & {[}\%{]} & {[}$ms${]} & up\\
\midrule

Ber-Mo & 98.4 & 71.6 & 0.168 & 0.843 & 2.99 & 13 \\
Ber-Tu & 98.5 & 58.4 & 0.129 & 1.029 & 3.01 & 12 \\
Ber-We & 98.6 & 61.3 & 0.141 & 1.249 & 2.97 & 13 \\
Ber-Th & 98.6 & 61.5 & 0.141 & 0.938 & 3.14 & 12 \\
Ber-Fr & 98.5 & 54.0 & 0.126 & 1.203 & 3.02 & 13 \\
Ber-Sa & 99.2 & 16.1 & 0.044 & 0.305 & 2.90 & 12 \\
Ber-Su & 99.3 & 12.9 & 0.032 & 0.700 & 2.50 & 14 \vspace{0.5em}\\
Ger-Mo & 95.8 & 122.1 & 0.177 & 0.869 & 5.90 & 104 \\
Ger-mid & 96.0 & 115.7 & 0.169 & 0.899 & 6.36 & 92 \\
Ger-Fr & 95.4 & 114.5 & 0.154 & 1.266 & 6.07 & 98 \\
Ger-Sa & 98.1 & 38.9 & 0.080 & 0.804 & 5.76 & 106 \\
Ger-Su & 98.6 & 17.0 & 0.037 & 0.379 & 5.63 & 108 \vspace{0.5em}\\
Eur-L & 53.2 & 2715.5 & 1.043 & 2.439 & 10.51 & 215 \\
Eur-M & 41.3 & 3219.8 & 1.084 & 3.450 & 8.87 & 257 \\
Eur-H & 40.2 & 3690.6 & 1.290 & 4.100 & 7.55 & 313 \\
\bottomrule
\end{tabular}
\end{center}

\caption{\label{tab:uniform-td-s+a}Random uniform earliest arrival
queries for TD-S+A.}
\end{table}

\begin{table}
\begin{centering}
\begin{tabular}{lrrrrrr}
\toprule
 & Run. & \multicolumn{5}{c}{Path Count}\\
\cmidrule(lr){3-7}
 & {[}ms{]} & avg. & Q90\% & Q99\% & Q99.9\% & max.\\
\midrule
Ber-Mo & 4.1 & 1.4 & 2 & 4 & 8 & 14\\
Ber-Tu & 4.1 & 1.4 & 2 & 4 & 7 & 12\\
Ber-We & 4.1 & 1.4 & 2 & 4 & 7 & 11\\
Ber-Th & 4.2 & 1.4 & 2 & 5 & 8 & 13\\
Ber-Fr & 4.0 & 1.4 & 2 & 4 & 8 & 15\\
Ber-Sa & 3.5 & 1.4 & 2 & 4 & 8 & 14\\
Ber-So & 3.4 & 1.4 & 2 & 4 & 6 & 11\vspace{0.5em}\\
Ger-Mo & 7.8 & 1.5 & 2 & 4 & 7 & 17\\
Ger-mid & 8.0 & 1.5 & 2 & 4 & 7 & 13\\
Ger-Fr & 7.5 & 1.4 & 2 & 4 & 7 & 12\\
Ger-Su & 6.3 & 1.3 & 2 & 4 & 6 & 13\\
Ger-So & 6.1 & 1.3 & 2 & 4 & 6 & 10\vspace{0.5em}\\
Eur-L & 16.7 & 62.3 & 102 & 124 & 133 & 140\\
Eur-M & 15.0 & 36.9 & 68 & 92 & 107 & 123\\
Eur-H & 14.0 & 23.2 & 47 & 72 & 90 & 111\\
\bottomrule
\end{tabular}
\par\end{centering}

\caption{\label{tab:profile}Running time and distinct path count for TD-S+P. The sampling rate $r$ is 10min.}
\end{table}

\section{Additional Dijkstra-Rank Plots}

Figures \ref{foo1} and \ref{foo2} present Dijkstra-rank plots analogues to \ref{fig:dij-ger-rank} for Ber-Tu and Eur-H.

\maketable{ber/di}{Dijkstra-Rank Results for Ber-Tu.}{foo1}
\maketable{eur/c3}{Dijkstra-Rank Results for Eur-H.}{foo2}


\newpage

\begin{thebibliography}{10}

\bibitem{adgw-arrn-13}
Ittai Abraham, Daniel Delling, Andrew~V. Goldberg, and Renato~F. Werneck.
\newblock Alternative routes in road networks.
\newblock {\em ACM Journal of Experimental Algorithmics}, 18(1):1--17, 2013.

\bibitem{bdgs-argrn-11}
Roland Bader, Jonathan Dees, Robert Geisberger, and Peter Sanders.
\newblock Alternative route graphs in road networks.
\newblock In {\em Proceedings of the 1st International ICST Conference on
  Theory and Practice of Algorithms in (Computer) Systems (TAPAS'11)}, volume
  6595 of {\em Lecture Notes in Computer Science}, pages 21--32. Springer,
  2011.

\bibitem{bdgmpsww-rptn-14}
Hannah Bast, Daniel Delling, Andrew~V. Goldberg, Matthias
  {M{\"u}ller--Hannemann}, Thomas Pajor, Peter Sanders, Dorothea Wagner, and
  Renato~F. Werneck.
\newblock Route planning in transportation networks.
\newblock Technical Report abs/1504.05140, ArXiv e-prints, 2016.
\newblock To appear in LNCS Volume on Algorithm Engineering, Lasse Kliemann and
  Peter Sanders (eds.).
\newblock URL: \url{http://arxiv.org/abs/1504.05140}.

\bibitem{bgsv-mtdtt-13}
Gernot~Veit Batz, Robert Geisberger, Peter Sanders, and Christian Vetter.
\newblock Minimum time-dependent travel times with contraction hierarchies.
\newblock {\em ACM Journal of Experimental Algorithmics}, 18(1.4):1--43, April
  2013.

\bibitem{bd-sharc-09}
Reinhard Bauer and Daniel Delling.
\newblock {SHARC}: Fast and robust unidirectional routing.
\newblock {\em ACM Journal of Experimental Algorithmics}, 14(2.4):1--29, August
  2009.
\newblock Special Section on Selected Papers from ALENEX 2008.
\newblock URL: \url{http://doi.acm.org/10.1145/1498698.1537599}.

\bibitem{bdpw-dtdrp-16}
Moritz Baum, Julian Dibbelt, Thomas Pajor, and Dorothea Wagner.
\newblock Dynamic time-dependent route planning in road networks with user
  preferences.
\newblock In {\em Proceedings of the 15th International Symposium on
  Experimental Algorithms (SEA'16)}, volume 9685 of {\em Lecture Notes in
  Computer Science}, pages 33--49. Springer, 2016.
\newblock URL:
  \url{http://link.springer.com/chapter/10.1007/978-3-319-38851-9_3}.

\bibitem{ch-tsrtn-66}
K.~Cooke and E.~Halsey.
\newblock The shortest route through a network with time-dependent internodal
  transit times.
\newblock {\em Journal of Mathematical Analysis and Applications},
  14(3):493--498, 1966.

\bibitem{d-tdsr-11}
Daniel Delling.
\newblock Time-dependent {SHARC}-routing.
\newblock {\em Algorithmica}, 60(1):60--94, May 2011.
\newblock URL: \url{http://dx.doi.org/10.1007/s00453-009-9341-0}.

\bibitem{dgpw-crprn-13}
Daniel Delling, Andrew~V. Goldberg, Thomas Pajor, and Renato~F. Werneck.
\newblock Customizable route planning in road networks.
\newblock {\em Transportation Science}, 51(2):566--591, 2017.
\newblock URL: \url{http://dx.doi.org/10.1287/trsc.2014.0579}.

\bibitem{dn-crdtd-12}
Daniel Delling and Giacomo Nannicini.
\newblock Core routing on dynamic time-dependent road networks.
\newblock {\em Informs Journal on Computing}, 24(2):187--201, 2012.

\bibitem{dw-lbrdg-07}
Daniel Delling and Dorothea Wagner.
\newblock Landmark-based routing in dynamic graphs.
\newblock In {\em Proceedings of the 6th Workshop on Experimental Algorithms
  (WEA'07)}, volume 4525 of {\em Lecture Notes in Computer Science}, pages
  52--65. Springer, June 2007.
\newblock URL:
  \url{http://www.springerlink.com/content/r18l738744xmn68v/?p=c3fb56d3275b4389bd305aef88ad40c8&pi=2}.

\bibitem{dw-tdrp-09}
Daniel Delling and Dorothea Wagner.
\newblock Time-dependent route planning.
\newblock In {\em Robust and Online Large-Scale Optimization}, volume 5868 of
  {\em Lecture Notes in Computer Science}, pages 207--230. Springer, 2009.

\bibitem{dgj-spndi-09}
Camil Demetrescu, Andrew~V. Goldberg, and David~S. Johnson, editors.
\newblock {\em The Shortest Path Problem: Ninth DIMACS Implementation
  Challenge}, volume~74 of {\em DIMACS Book}.
\newblock American Mathematical Society, 2009.

\bibitem{dbs-a-10}
Ugur Demiryurek, Farnoush Banaei-Kashani, and Cyrus Shahabi.
\newblock A case for time-dependent shortest path computation in spatial
  networks.
\newblock In {\em Proceedings of the 18th ACM SIGSPATIAL International
  Conference on Advances in Geographic Information Systems (GIS'10)}, pages
  474--477, 2010.

\bibitem{dsw-cch-15}
Julian Dibbelt, Ben Strasser, and Dorothea Wagner.
\newblock Customizable contraction hierarchies.
\newblock {\em ACM Journal of Experimental Algorithmics}, 21(1):1.5:1--1.5:49,
  April 2016.
\newblock URL: \url{http://doi.acm.org/10.1145/2886843}.

\bibitem{d-ntpcg-59}
Edsger~W. Dijkstra.
\newblock A note on two problems in connexion with graphs.
\newblock {\em Numerische Mathematik}, 1(1):269--271, 1959.

\bibitem{d-aassp-69}
Stuart~E. Dreyfus.
\newblock An appraisal of some shortest-path algorithms.
\newblock {\em Operations Research}, 17(3):395--412, 1969.

\bibitem{fhs-octds-14}
Luca Foschini, John Hershberger, and Subhash Suri.
\newblock On the complexity of time-dependent shortest paths.
\newblock {\em Algorithmica}, 68(4):1075--1097, April 2014.

\bibitem{gssv-erlrn-12}
Robert Geisberger, Peter Sanders, Dominik Schultes, and Christian Vetter.
\newblock Exact routing in large road networks using contraction hierarchies.
\newblock {\em Transportation Science}, 46(3):388--404, August 2012.

\bibitem{gh-cspas-05}
Andrew~V. Goldberg and Chris Harrelson.
\newblock Computing the shortest path: {A*} search meets graph theory.
\newblock In {\em Proceedings of the 16th Annual {ACM--SIAM} Symposium on
  Discrete Algorithms (SODA'05)}, pages 156--165. SIAM, 2005.

\bibitem{hsw-emlog-08}
Martin Holzer, Frank Schulz, and Dorothea Wagner.
\newblock Engineering multilevel overlay graphs for shortest-path queries.
\newblock {\em ACM Journal of Experimental Algorithmics}, 13(2.5):1--26,
  December 2008.

\bibitem{k-hdara-13}
Moritz Kobitzsch.
\newblock {HiDAR}: An alternative approach to alternative routes.
\newblock In {\em Proceedings of the 21st Annual European Symposium on
  Algorithms (ESA'13)}, volume 8125 of {\em Lecture Notes in Computer Science},
  pages 613--624. Springer, 2013.

\bibitem{krs-eepma-13}
Moritz Kobitzsch, Marcel Radermacher, and Dennis Schieferdecker.
\newblock Evolution and evaluation of the penalty method for alternative
  graphs.
\newblock In {\em Proceedings of the 13th Workshop on Algorithmic Approaches
  for Transportation Modeling, Optimization, and Systems (ATMOS'13)},
  OpenAccess Series in Informatics (OASIcs), pages 94--107, 2013.
\newblock URL: \url{http://drops.dagstuhl.de/opus/volltexte/2013/4247}.

\bibitem{kmppwz-eotdr-16}
Spyros Kontogiannis, George Michalopoulos, Georgia Papastavrou, Andreas
  Paraskevopoulos, Dorothea Wagner, and Christos Zaroliagis.
\newblock Engineering oracles for time-dependent road networks.
\newblock In {\em Proceedings of the 18th Meeting on Algorithm Engineering and
  Experiments (ALENEX'16)}. SIAM, 2016.
\newblock URL: \url{http://epubs.siam.org/doi/abs/10.1137/1.9781611974317.1}.

\bibitem{l-aefea-04}
Ulrich Lauther.
\newblock An extremely fast, exact algorithm for finding shortest paths in
  static networks with geographical background.
\newblock In {\em Geoinformation und Mobilit{\"a}t - von der Forschung zur
  praktischen Anwendung}, volume~22, pages 219--230. IfGI prints, 2004.

\bibitem{ls-csarr-13}
Dennis Luxen and Dennis Schieferdecker.
\newblock Candidate sets for alternative routes in road networks.
\newblock {\em ACM Journal of Experimental Algorithmics}, 2013.
\newblock Submitted to.

\bibitem{ndls-bastd-12}
Giacomo Nannicini, Daniel Delling, Leo Liberti, and Dominik Schultes.
\newblock Bidirectional {A*} search on time-dependent road networks.
\newblock {\em Networks}, 59:240--251, 2012.
\newblock Best Paper Award.

\bibitem{or-spmda-90}
Ariel Orda and Raphael Rom.
\newblock Shortest-path and minimum delay algorithms in networks with
  time-dependent edge-length.
\newblock {\em Journal of the ACM}, 37(3):607--625, 1990.

\bibitem{sww-daola-00}
Frank Schulz, Dorothea Wagner, and Karsten Weihe.
\newblock {D}ijkstra's algorithm on-line: An empirical case study from public
  railroad transport.
\newblock {\em ACM Journal of Experimental Algorithmics}, 5(12):1--23, 2000.

\end{thebibliography}
\end{document}